\input harvmac
\noblackbox

\def\bfone{\relax{\rm 1\kern-.35em 1}}
\def\inbar{\vrule height1.5ex width.4pt depth0pt}
\def\LG{Lan\-dau-Ginz\-burg\ }
\def\IC{\relax\,\hbox{$\inbar\kern-.3em{\mss C}$}}
\def\ID{\relax{\rm I\kern-.18em D}}
\def\IF{\relax{\rm I\kern-.18em F}}
\def\IH{\relax{\rm I\kern-.18em H}}
\def\II{\relax{\rm I\kern-.17em I}}
\def\IN{\relax{\rm I\kern-.18em N}}
\def\IP{\relax{\rm I\kern-.18em P}}
\def\IQ{\relax\,\hbox{$\inbar\kern-.3em{\rm Q}$}}
\def\us#1{\underline{#1}}
\def\IR{\relax{\rm I\kern-.18em R}}
\font\cmss=cmss10 \font\cmsss=cmss10 at 7pt
\def\ZZ{\relax\ifmmode\mathchoice
{\hbox{\cmss Z\kern-.4em Z}}{\hbox{\cmss Z\kern-.4em Z}}
{\lower.9pt\hbox{\cmsss Z\kern-.4em Z}}
{\lower1.2pt\hbox{\cmsss Z\kern-.4em Z}}\else{\cmss Z\kern-.4em
Z}\fi}

\def\cP{{\cal P}} 
\def\cR{{\cal R}} 
\def\nup#1({Nucl.\ Phys.\ $\us {B#1}$\ (}
\def\plt#1({Phys.\ Lett.\ $\us  {B#1}$\ (}
\def\cmp#1({Comm.\ Math.\ Phys.\ $\us  {#1}$\ (}
\def\prp#1({Phys.\ Rep.\ $\us  {#1}$\ (}
\def\prl#1({Phys.\ Rev.\ Lett.\ $\us  {#1}$\ (}
\def\prv#1({Phys.\ Rev.\ $\us  {#1}$\ (}
\def\mpl#1({Mod.\ Phys.\ Let.\ $\us  {A#1}$\ (}
\def\ijmp#1({Int.\ J.\ Mod.\ Phys.\ $\us{A#1}$\ (}
\def\jag#1({Jour.\ Alg.\ Geom.\ $\us {#1}$\ (}
\def\tit#1|{{\it #1},\ }
\def\Coeff#1#2{{#1\over #2}}
\def\Coe#1.#2.{{#1\over #2}}
\def\coeff#1#2{\relax{\textstyle {#1 \over #2}}\displaystyle}
\def\coe#1.#2.{\relax{\textstyle {#1 \over #2}}\displaystyle}

\def\doubref#1#2{\refs{{#1},{#2}}}

\def\br{\hfill\break}
%
%
\lref\SW{N.\ Seiberg and E.\ Witten, \nup426(1994) 19,
hep-th/9407087; \nup431(1994) 484, hep-th/9408099.}
\lref\Kanev{V.\ Kanev, {\it Spectral curves, simple Lie algebras,
and Prym-Tjurin varieties}, in {\sl Theta Functions, Bowdoin 1987},
Proc. Symp. Pure Math. 49,
L.\ Ehrenpreis and R.\ Gunning (eds.), AMS, Providence (1989).}
\lref\Donagi{R.\ Donagi, {\it Decomposition of spectral covers}, in
{\sl Journ\'ees de Geometrie Alg\'ebrique d'Orsay}, Asterisque vol.
218,Soc. Math. de France, Paris (1993).}
\lref\MW{E.\ Martinec and N.P. \ Warner, \nup{459} (1996) 97,
hep-th/9509161}
\lref\Dubr{B.Dubrovin, {\it Geometry of $2d$ Topological Field
Theories},
     SISSA-89/94/FM; \br
   {\it  Integrable Systems in Topological Field Theory},
     \nup {379} (1992) 627-689; \br
   {\it Hamiltonian Formalism of
     Witham-type Hierarchies in Topological
     Landau-Ginzburg Model}, Comm.\ Math.\ Phys. {\bf 145} (1992)
19.}
\lref\DVV{R.\ Dijkgraaf, E. Verlinde and H. Verlinde, \nup{352}
(1991)
59.}
\lref\WLNW{W.\ Lerche and N.P.\ Warner, \nup358 (1991) 571.}
\lref\KLMVW{A. \ Klemm, W.\ Lerche, P. \ Mayr, C. \ Vafa and
N.P. \ Warner, hep-th/9604034.}
\lref\KV{S.\ Kachru and C.\ Vafa, \nup450 (1995) 69, hep-th/9505105.}
\lref\KKLMV{S.\ Kachru, A.\ Klemm, W.\ Lerche, P.\ Mayr and
C.\ Vafa, \nup459 (1996) 537, hep-th/9508155.}
\lref\RHarts{R.Hartshorne, {\it Algebraic Geometry},
Springer-Verlag (1977).}
\lref\SKDM{S.\ Katz and D.\ Morrison, \jag{1} (1992) 449.}
\lref\JMDN{J.\ Minahan and D.\ Nemeschansky, hep-th/9608047.}
\lref\Arn{V.\ Arnold, A.\ Gusein-Zade and A.\
Varchenko, {\it Singularities of Differentiable Maps II},
Birkh\"auser 1985.}
\lref\strom{A.\ Strominger, \nup451 (1995) 96, hep-th/9504090.}
\lref\KLTY{A.\ Klemm, W.\ Lerche, S.\ Theisen and S.\ Yankielowicz,
\plt344(1995) 169, hep-th/9411048.}
\lref\AF{P. Argyres and A. Faraggi, {Phys.\ Rev.\ Lett.}
     {\bf 74} (1995) 3931, hep-th/9411057.}
\lref\DS{U.\ Danielsson and B.\ Sundborg, \plt B358 (1995)
273, hep-th/9504102.}
\lref\BL{A.\ Brandhuber and K.\ Landsteiner,
\plt B358 (1995) 73, hep-th/9507008.}
\lref\sdstrings{M.\ Duff and J.\ Lu, \nup416 (1994) 301,
hep-th/9306052; \br E.\ Witten,
preprint IASSNS-HEP-95-63, hep-th/9507121;\br
A.\ Strominger, hep-th/9512059; \br
O.\ Ganor and A.\ Hanany, preprint IASSNS-HEP-96-12,
hep-th/9602120;\br N.\ Seiberg and E.\ Witten, preprint RU-96-12,
hep-th/9603003;\br M.\ Duff, H.\ Lu and C.N.\ Pope, preprint
CTP-TAMU-9-96, hep-th/9603037.}
\lref\russ{A.\ Gorskii, I.\ Krichever, A.\ Marshakov, A.\ Mironov
and  A.\ Morozov, \plt{355} (1995) 466, hep-th/9505035;\br
H.\ Itoyama and, A.\ Morozov, preprints ITEP-M5-95 and ITEP-M6-95,
hep-th/9511126 and hep-th/9512161.}
\lref\TEKH{T.\ Eguchi and K.\ Hori, preprint UT-755, hep-th/9607125.}
\lref\BDS{T.\ Banks, M.\ Douglas and  N.\ Seiberg,
preprint RU-96-41, hep-th/9605199.}
\lref\AD{P.\ Argyres and M.\ Douglas, \nup448 (1995) 93, hep-th/9505062.}
%
%
\Title{\vbox{
\hbox{CERN-TH/96-237}
\hbox{USC-96/022}
\hbox{\tt hep-th/9608183}
}}{Exceptional SW Geometry from ALE Fibrations}

\bigskip
\centerline{W.~Lerche}
\bigskip\centerline{{\it Theory Division, CERN, 1211 Geneva 23,
Switzerland}}
\bigskip
\centerline{and}
\bigskip
\centerline{N.P.~Warner}
\bigskip
\centerline{{\it Physics Department, U.S.C., University Park,
Los Angeles, CA 90089}}

\vskip .3in

We show that the genus 34 Seiberg-Witten curve underlying $N=2$
Yang-Mills theory with gauge group $E_6$ yields physically equivalent
results to the manifold obtained by fibration of the $E_6$ ALE
singularity. This reconciles a puzzle raised by $N=2$ string duality.

\vskip .3in


\Date{\vbox{\hbox{CERN-TH/96-237}\hbox{\sl {August 1996}}}}

%
\parskip=4pt plus 15pt minus 1pt
\baselineskip=15pt plus 2pt minus 1pt
%
\newsec{Introduction}

There is now an extensive body of literature on the construction of
the quantum effective actions on the Coulomb branch of $N=2$
supersymmetric Yang-Mills theories \SW. The work of \SW\ was extended
to other gauge groups: for example, to $G=SU(n)$ in \doubref\KLTY\AF,
to
$G=SO(2n+1)$ in \DS, and to $G=SO(2n)$ in \BL. In these approaches,
the curves are given in terms of the appropriate simple singularities
$W_{ADE}$ \Arn, and are generically of the form
\eqn\ssform{y^2\ = W^2(x; u_j)-\mu^{2}\ ,}
with $\mu=\Lambda^{h^\vee}$, where $h^\vee$ is the dual Coxeter
number of $G$, and $\Lambda$ is the quantum scale. The $u_j,
j=1,\dots,\ell$ are the fundamental Casimir invariants (with degree
increasing with the subscript $j$) and $\ell$ is the rank of $G$; the
top Casimir, $u_\ell$, has degree $h^\vee$. For example, for
G=$SU(n)$, one has $W_{A_{n-1}}(x; u_j)=x^n-\sum_{j=1}^{n-1}
u_j x^{n-1-j}$.

In a complementary, unifying approach based on integrable
systems \russ, a general scheme for obtaining Seiberg-Witten 
(SW) curves for all
groups was presented in \MW. As explained in more detail below,
these curves are of the form
\eqn\mwform{\zeta + {\mu^2\over\zeta} + P_{\cal R} (x; u_j)\ =\ 0\ ,}
where $P_{\cal R}$ is a polynomial in $x$ of order $dim({\cal R})$,
where $\cal R$ is some representation of $G$. For $G=SU(n)$, one can
take $P_{\cal R}(x; u_j)\equiv W_{A_{n-1}}(x; u_j)$ so that the
curves \mwform\ and \ssform\ are manifestly the same, up to a simple
reparametrization. This however cannot be done for other groups,
since
$dim({\cal R})$ will in general\foot {However, this can easily be
reconciled \BL\ for $G=SO(2n)$.} not match the degree of $W_{ADE}$,
which is equal to ${h^\vee}$.

More recently, it was found in \KLMVW\ how, via $N=2$ heterotic-type
II string duality \KV, local SW geometry can be derived from
fibrations of ALE spaces: the relevant manifolds are described by
\eqn\ALEform{\zeta + {\mu^2\over\zeta} +
 W^{{\rm ALE}}_{ADE}(x_1,x_2,x_3; u_j)\ =\ 0\ ,}
where $W^{ALE}_{ADE}$ is the (non-compact) ALE space of type $ADE$;
for G=$SU(n)$, $W^{ALE}_{A_{n-1}} \equiv W_{A_{n-1}} (x_1; u_j)+
{x_2}^2+{x_3}^2$. Obviously, by trivially integrating out the
quadratic pieces in $x_2$ and $x_3$ (which does not change the
singularity type) this manifold is equivalent to the above SW curves.
It also gives rise to the same periods \KLMVW.

Actually, by simple reparametrization, \ALEform\ can be brought into
the form \ssform\ (but with $W$ depending on more than one variable) 
for all gauge groups, and it was conjectured in \KLTY\ that this
should describe the Seiberg-Witten effective action.  
String duality implies that this must indeed be true. However, for 
the groups $G=E_n, n \ge 6$, $x_2$ does not enter quadratically and 
thus cannot easily be integrated out (for example, one has $W^{{\rm
ALE}}_{E_6}={x_1}^3+{x_2}^4+{x_3}^2+...$). This means that for $E_n$,
the manifolds \ALEform\ have {\it a priori} no obvious relation to
Riemann surfaces, and appear intrinsically as higher dimensional
surfaces.

The question then immediately arises as to how the manifolds
$\ALEform$ are related to the curves \mwform\ for exceptional gauge
groups.  It is very natural to believe
that somehow the periods must be the same, but how this
precisely works was not clear until now. It is the purpose of this
letter to show that from the point of view of SW theory, the curves
\mwform\ and the manifolds \ALEform\ are indeed
physically equivalent.

There is also a physics aspect to this. In fact,
\ALEform\ represents (a local, non-compact piece of) a
threefold, and the BPS states correspond to wrappings of
type IIB 3-branes \strom\ around the local vanishing homology $H_3$.
For $G=SU(n)$, it follows from the considerations in \KLMVW\
that upon integrating out the quadratic terms in $x_{2}$ and 
$x_{3}$, the wrapped 3-branes are {\it effectively} equivalent to
wrapping 1-branes around the curve \mwform; the 1-branes,
which are the left-over pieces of the 3-branes, are precisely
the non-criticial, (anti-)self dual strings of \sdstrings.
On physical grounds, one would expect this to be true also for the 
other, and in particular the exceptional, gauge groups, but
this is at first sight not so obvious. By performing the integrals
over $x_2$ and $x_3$ in $W^{{\rm ALE}}_{E_6}(x_1,x_2,x_3; u_j)$,
we will argue that indeed the 3-branes, when wrapped around
the vanishing cycles of \ALEform, are equivalent to 1-branes
on the curves \mwform.
We expect that a similar story holds for the
other exceptional groups as well.

The potential importance of this type of considerations lies not in
the generalization to other gauge groups of what has been known
before, but ultimately in the generalization to other, non-field
theory limits of type IIB strings. In a threefold, more general
singularities than those describing YM theories typically do arise,
and the question is under what conditions $\alpha'\to 0$ physics can
effectively be described in terms of SW-like curves (with
anti-self-dual strings wrapping around them).


\newsec{Seiberg--Witten Curves}

In \MW\ a general scheme was presented for obtaining a
family of Seiberg-Witten Riemann surfaces for an $N=2$
supersymmetric Yang-Mills theory with arbitrary gauge
group, $G$.  Specifically, given any representation ${\cal R}$
of $G$, one considers the polynomial:
\eqn\pdefn{P_{\cal R} (x; u_j) ~=~ \det(x ~-~ \Phi_0) \ ,}
where $\Phi_0$ is a generic adjoint Higgs v.e.v. written in the
representation ${\cal R}$.
One can diagonalize $\Phi_0$ to some Cartan subalgebra element
$v \cdot H$, where $v$ is an $\ell$-dimensional vector.  In the
representation $\cR$, the eigenvalues of $\Phi_0$ are $\lambda
\cdot v$, where $\lambda$ are the weights of $\cR$.  Thus one
can write:
\eqn\Pasprod{P_\cR ~=~ \prod_\lambda ~(x ~-~ \lambda \cdot v) \ .}
The polynomial, $P_\cR (x; u_j)$, naturally decomposes into
factors corresponding to Weyl orbits of weights.  If the
representation ${\cal R}$ is miniscule, then by definition,
the weights of ${\cal R}$ form exactly one such orbit.  If the
representation is not miniscule, then we will take $P_\cR$
to be any one of the Weyl orbit factors in the determinant
\pdefn.  Let $M$ be the degree of $P_\cR$.

To get the Riemann surface for a simply laced group, $G$, one
first defines a function $\tilde P_\cR (x,\zeta;u_j, \mu)$ via
\eqn\ptilde{\tilde P_\cR (x,\zeta;u_j, \mu) ~\equiv~ P_\cR
(x;u_1, \dots, u_{\ell-1}, u_\ell + \zeta + \mu^2/\zeta) \ .}
That is, one shifts the top Casimir by $u_\ell \to
u_\ell + \zeta + \mu^2/\zeta$.  The Riemann surface is then
given by
\eqn\Rsurf{\tilde P_\cR (x,\zeta;u_j, \mu) ~=~ 0}
The canonical way to view this surface
\refs{\Kanev,\Donagi,\MW} is as an $M$-sheeted foliation by
$x(\zeta)$ over the base $\zeta$-sphere.  The sheets are
then in one-to-one correspondence with the weights of $\cR$
(or at least the Weyl orbit of weights that one has chosen).
There are $2(\ell + 1)$ branch points in the base, and they come
in pairs related by the involution symmetry $\zeta \to \mu^2/\zeta$.
The first $\ell$ pairs of branch points can be labelled by
a system of simple roots $\{\alpha_j, j = 1,\dots,\ell\}$ of
$G$, and the monodromy around the branch point is then
given by the Weyl reflection $r_{\alpha_j}$.
Above the $\alpha_j$-cut one then connects sheets in pairs
according to whether the weights labelling the sheets are
exchanged by the Weyl reflection $r_{\alpha_j}$.  The
$(\ell +1)$-th pair of branch points is $\{\zeta=0,
\zeta = \infty\}$ and each of these points has the Coxeter
monodromy corresponding to the product of the fundamental
Weyl reflections.  The sheets are joined above $\zeta =0$
and $\zeta = \infty$ according to the Coxeter orbits of the
weight labels.

The genus of the Riemann surface is usually far larger than the
rank, $\ell$, of $G$.  Thus one needs to isolate a special
sub-Jacobian, or, more precisely, a special Prym Variety, whose
periods give the effective action of the $U(1)^\ell$ on the
Coulomb branch.  This can be done using the underlying
integrable system, and can be implemented directly
in a number of ways \refs{\Kanev,\Donagi,\MW}.  The simplest is
to first take the $\ell$ cycles that surround the $\alpha_j$-cuts
on the $\zeta$-sphere, and then take the inverse image under the
projection of the Riemann surface to the $\zeta$-sphere.  This gives
$\ell$ $A$-cycles.  The $B$-cycles are then obtained by finding the
cycles that intersect the $A$-cycles in the proper manner.
As we will see later, once one has the Seiberg-Witten differential
$\lambda_{SW}$, the issue of the proper $A$ and $B$ cycles
is essentially moot.

Because of the connection with integrable systems, the Seiberg-Witten
differential takes the universal form:
\eqn\lamSW{\lambda_{SW} ~=~ -2 x ~{d\zeta \over \zeta} \ .}

We now focus on the details of the curve for $E_6$.  The simplest
$E_6$ curve is obtained from the (miniscule) $27$-dimensional
representation.  Given a system of simple roots, $\alpha_1,
\dots, \alpha_6$, one finds that for each root, $\alpha_j$,
there are six\foot{The fact that this is the rank of $E_6$ is
a coincidence.}  weights, $\lambda_j^{(a)}$, of the $27$ such that
 $\lambda_j^{(a)} \cdot \alpha_j = + 1$.  Consequently,
 $\lambda_j^{\prime (a)} \equiv  \lambda_j^{(a)} - \alpha_j$
are weights of the $27$ with  $\lambda_j^{\prime (a)} \cdot
\alpha_j = - 1$.  Thus above {\it each} of the six $\alpha_j$-cuts
on the $\zeta$-sphere, the sheets of the foliation are connected in
six pairs, making a total of 36 such interconnections.  Under the
action of the Coxeter element the weights form three orbits of
order $12$, $12$ and $3$ respectively.  If one imagines
assembling the surface by first making the connections at
$\zeta=0$ and $\zeta=\infty$, one first gets three disjoint spheres
(the Coxeter orbits) that must then be laced together by the
$36$ pieces of plumbing mentioned earlier.  The result is a
genus $34$ surface.

This surface is explicitly given by
\eqn\Esixsurf{\coeff{1}{2}~ x^3 ~\tau^2 ~-~q_1~\tau ~+~ q_2 ~=~ 0 \ ;
\qquad \tau ~\equiv~ \zeta ~+~ {\mu^2 \over \zeta} ~+~ u_6\ , }
where
\eqn\ponetwo{\eqalign{ q_1 ~=&~   270~x^{15} + 342 ~u_1~x^{13} +
162~u_1^2~x^{11} - 252~u_2~x^{10}  +  (26~u_1^3 + 18~u_3)~x^9 \cr
&~- 162~u_1 u_2~x^8 + (6~u_1 u_3 - 27~u_4)~x^7
 - (30~u_1^2 u_2  - 36~u_5)~x^6 \cr
&~ + (27~u_2^2 - 9~u_1 u_4)~x^5 -  (3~u_2 u_3 - 6~u_1 u_5) ~x^4
- 3~u_1 u_2^2~x^3\cr
&~ - 3~u_2 u_5~x  - u_2^3 \ ;\cr
q_2 ~=&~ {1 \over 2 x^3}~ (q_1^2 ~-~ p_1^2~p_2) \ ; \cr
p_1 ~=&~  78~x^{10} + 60~u_1 ~x^8 + 14~u_1^2~x^6  - 33~u_2~x^5
 + 2~u_3 ~x^4 - 5~u_1 u_2~x^3 - u_4~x^2 \cr
&~ - u_5~x - u_2^2\ ; \cr
p_2 ~=&~ 12~x^{10} + 12~u_1~x^8 + 4u_1^2~x^6 - 12~u_2~x^5
+ u_3~x^4 - 4~u_1 u_2~x^3  - 2~u_4~x^2 \cr
&~ + 4~u_5~x  + u_2^2 \ .}}
Note that \Esixsurf\ is {\it not} a hyperelliptic curve, unlike the
simplest curves for the $A_n$ and $D_n$ groups. Moreover, like the
curves for $D_n$, the genus of the curve exceeds the rank of the
gauge group, but unlike $D_n$, there is no obvious, elementary
symmetry that picks out the cycles that yield the quantum effective
action \BL. For the curve defined by \Esixsurf\ one has to use the
methods outlined above and in \MW\ to determine the cycles of
interest.

It is useful to note that since
\Esixsurf\ is quadratic in $\tau$, one can solve it to obtain a
more convenient presentation of the curve:
\eqn\EsixW{\tau ~=~ \zeta ~+~ \mu^2/\zeta ~+~ u_6 ~=~ {1 \over x^3}~
\big[~ q_1 ~\pm~ p_1 ~\sqrt{p_2}~\big] \ , }
where $q_1, p_1$ and $ p_2$ are defined above.

This expression for $\tau$ will be of importance in the next section,
but it is also of interest for other reasons. One may view \EsixW\ as
determining how each eigenvalue, $x$, of the v.e.v., $\Phi_0$,
changes as one varies the top Casimir, $u_6$, while holding the
remaining Casimirs, $u_1, \dots, u_5$ fixed. As a result, the
explicit expression \EsixW, which has degree $12$, can be interpreted
as a single variable (though non-polynomial) version of the \LG
potential for $E_6$, in the sense of \Dubr. It can probably be
used to describe the coupling of $E_6$ minimal matter to topological
gravity directly, using the residue methods of
\refs{\DVV,\Dubr}\foot{Indeed, there was a long-standing question how
the KdV type Lax operator, which is derived from the 27-dimensional
representation of $E_6$ via Drinfeld-Sokolov reduction and which thus
has {\it a priori} degree 27, is related to the $E_6$ simple
singularity of degree 12. This simple singularity supposedly figures
as the superpotential of a topological LG theory that describes the
matter-gravity system \DVV.}.

Finally, we observe that one can also obtain \EsixW\
from decomposing Casimir invariants invariants of $E_6$ into those
of $SO(10) \times U(1)$. If one writes the $u_j$ in terms of the
five invariants of $SO(10)$ and the $U(1)$ eigenvalue $x$, and
then inverts this relationship, one can easily extract an expression
for $u_6$ in terms of $u_1, \dots, u_5$ and $x$.  Thus one can
easily derive \EsixW\ from the results of \WLNW.
That this is equivalent to the procedure described above follows
from the fact that the stabilizer of a weight space in
the $27$ of $E_6$ is $SO(10) \times U(1)$.

\newsec{The quantum effective action from string theory}

It was argued in \refs{\KV, \KKLMV, \KLMVW} that one can obtain the
quantum effective action of a pure gauge theory from the IIB string
compactified on a Calabi-Yau manifold that is degenerating to a
nearly singular ALE fibration over a $\IP^1$ base. The Riemann
surface can then be constructed\foot{Note that, strictly speaking,
the SW curve itself is not geometrically embedded in the Calabi-Yau
manifold.} by using the monodromy data on the non trivial $2$-cycles
of the fiber to produce the monodromies of the Riemann sheets. The
Seiberg-Witten differential is obtained by integrating the
holomorphic $(3,0)$-form, $\Omega$, of the Calabi-Yau manifold over
the $2$-cycles in the fiber, yielding a meromorphic $1$-form on the
Riemann surface \KLMVW. This is easily verified for $A_n$ gauge
groups,\foot{It is straightforward to extend the arguments of \KLMVW\
to $D_n$ gauge groups as well.} but it is far less obvious for the
exceptional groups. Here we will describe, in detail, how is works
for $E_6$.

The singular ALE fibration for $E_6$ has the form:
\eqn\ALE{{\cal P} (y_1,y_2,y_3; \zeta) ~\equiv~ W_{E_6}
(y_1,y_2,y_3) ~+~\nu~(\zeta ~+~ \mu^2/\zeta) ~=~ 0\ ,}
where $\zeta$ is the coordinate on the base $\IP^1$ and
\eqn\EsixWnew{\eqalign{W_{E_6}(y_1,y_2,y_3) ~=~ y_1^3 ~+~
y_2^4 ~+~ y_3^2 &~+~ w_1~y_1~y_2^2 ~+~  w_2~y_1~y_2 ~+~
w_3~y_2^2 \cr &~+~ w_4~y_1 ~+~ w_5~y_2   ~+~ w_6 \ .}}
The parameter $\nu$ in \ALE\ is a normalization constant
that has been inserted for later convenience.
The holomorphic $(3,0)$-form in these local coordinates can be
written:
\eqn\Omform{\Omega ~=~ \bigg( {d \zeta \over \zeta}\bigg)~\wedge~
{dy_1 \wedge dy_2 \over y_3} ~=~ \bigg( {d \zeta \over \zeta}
\bigg)~\wedge~ {dy_1 \wedge dy_2 \over \sqrt{{\cal P}
(y_1,y_2,y_3 = 0; \zeta)}} \ .}
The appearance of $d \zeta/\zeta$ in \Omform\ follows from the
Ricci flatness of the underlying Calabi-Yau and the
Ricci flatness of the fiber.

We need to find the $2$-cycles in the fiber space defined by
$W_{E_6} = const$, and then we need to integrate \Omform\ over
these fibers.  The key to doing this is to recall some beautiful
facts of classical algebraic geometry \RHarts.  One starts by
recasting \EsixWnew\ as a cubic in $\IP^3$ \SKDM.  That is, given
\eqn\cubic{\eqalign{W(x_i) ~=~ x_3^2~x_4 ~+~ x_1^3 ~+~
2 i~x_2^2~x_3 & ~+~  w_1~x_1~x_2^2 ~+~ w_2~x_1~x_2~x_4 ~+~
w_3~x_2^2~x_3 \cr & ~+~ w_4~x_1~x_4^2 ~+~
w_5~x_2~x_4^2   ~+~ w_6~x_4^3\ ,} }
in homogeneous coordinates, $x_i$, one can obtain \EsixWnew\ by
going to the patch $x_4 \ne 0$, setting  $ y_1 =x_1/x_4$,
$y_2 = x_2/x_4$ and $y_3 = (x_3/x_4 + i x_2^2/x_4^2)$.
A cubic in $\IP^3$ can be thought of as the blow-up
of six points in general position in $\IP^2$ \RHarts.
The second homology, $H_2$, of this surface is seven dimensional,
and consists of these six (non-intersecting) spheres, and the
canonical class of the $\IP^2$, which has intersection number
$1$ with each of the six blown up spheres.  The six
dimensional integer homology of the local $E_6$
singularity consists on a six-dimensional subspace of $H_2$.
This subspace is obtained by orthogonalizing with
respect to a vector with coordinates $(3,-1,-1,-1,-1,-1,-1)$,
and the result is a set of homology cycles that have
the intersection matrix of $E_6$ \refs{\RHarts,\SKDM,\Kanev}.

Thus far, we have seen very little that looks like the
$27$-sheeted Riemann surface described in the previous
section.  However, a celebrated fact about a cubic in
$\IP^3$ is that it contains $27$ lines.  That is, there
are $27$ holomorphically embedded $\IP^1$'s in this space.
Each of these $27$ lines is a non-trivial element of homology,
and together they (over) span the (seven dimensional) homology.
Thus, we can
exhibit the homology by exhibiting these lines.  Moreover,
it is also well know that the monodromy group of the $E_6$
singularity is the Weyl group of $E_6$, and that it acts on
these lines as on the $27$ of $E_6$ \foot{As a representation
of the Weyl group of $E_6$, the $27$ is reducible:
$27 = 20 + 6 +1$.  The singlet is the cycle with respect
to which one orthogonalizes to get the cycles of the $E_6$
singularity, and the $6$ is the fundamental reflection
representation  (the Cartan subalgebra representation) on
the compact homology basis of the $E_6$ singularity.}.

The explicit computation of the $27$ lines for the generic
$E_6$ singularity was recently given by Minahan and Nemeschansky
\JMDN, and while the motivation of these authors was rather
different from ours, our analysis will closely parallel theirs.
Make a change of variables from $(y_1,y_2)$ to $(x,y)$ in
\EsixWnew, where $y_1 = x~y + \alpha(x)$, and
$y_2 =  y + \beta(x)$. As yet, $\alpha$ and $\beta$ are
arbitrary functions of $x$.  With these changes of
variable, the function ${\cal P}$ in \ALE\ is a quartic
in $y$.  Choosing $\beta = -(x^3 + w_1 x)/4$
cancels the $y^3$ term in this quartic.  One next chooses
$\alpha(x)$ so as to get rid of the linear term in $y$.
This involves solving a quadratic equation for $\alpha(x)$,
however, before doing this, we wish to reparametrize the
versal deformation of the $E_6$ singularity via:
\eqn\wintou{\eqalign{w_1 ~=~& \coeff{1}{2}~u_1 \ ; \qquad
w_2 ~=~ - \coeff{1}{4}~u_2 \ ; \qquad w_3 ~=~ \coeff{1}{96}~
(u_3 - u_1^3) \ ; \cr
w_4 ~=~& \coeff{1}{96}~(u_4 + \coeff{1}{4} u_1~u_3 - \coeff{1}{8}
u_1^4) \ ; \qquad
w_5 ~=~  - \coeff{1}{48}~(u_5 - \coeff{1}{4} u_1^2 u_2) \ ; \cr
w_6 ~=~& \coeff{1}{3456}(u_6 ~+~ \coeff{1}{16}~u_1^6 -
\coeff{3}{16}~u_1^3~u_3 + \coeff{3}{32} u_3^2 -
\coeff{3}{4}~u_1^2~u_4)\ .}}
One then finds that
\eqn\alphx{ \alpha(x) ~=~ {1 \over 48 x}~ \big[~(2 u_1~x^3 +
u_1^2~x + 2u_2) ~\pm~ 2~\sqrt{p_2}~\big] \ .}
where $p_2$ is given by \ponetwo.

The lines in cubic occur when one
can analytically solve ${\cal P} = 0$.  Since ${\cal P}$
is quadratic in $y_3$, this happens precisely when the
rest of $\cP$ is a perfect square.  That is, we must
find the points at which $\cP(y_1, y_2, y_3 = 0; \zeta)$
becomes a perfect square.  With the changes of variable
above, and the choices of $\alpha(x)$ and $\beta(x)$,
the function ${\cal P}(x,y,y_3 = 0)$ takes the form of a
quadratic in $y^2$.  The lines are thus defined by the
vanishing of the discriminant, $\Delta$, of this quadratic
in $y^2$. A straightforward computation shows that:
\eqn\discrim{\Delta ~=~ \Coeff{1}{864}~ \Big[~{{q_1 ~\mp~ p_1~
\sqrt{p_2}} \over x^3} ~-~ u_6~\Big] ~-~ 4~\nu~(\zeta ~+~ \mu^2/
\zeta) \ .}
Taking $\nu = {1 \over 3456}$, one thus finds that is discriminant
vanishes precisely when $x$ and $\zeta$ satisfy the equation
\Esixsurf.  We thus see that the Riemann surface is nothing other
than a fibration, over the $\zeta$-sphere, of the locations of
the $27$ lines in the $E_6$ singularity.

The lines are all defined by a relationship of the form:
\eqn\lines{y_3^2 ~=~ - (y^2 ~+~ a(x))^2 \ ,}
and hence, from the perspective of the singularity polynomial,
these lines are all non-compact cycles in the closed homology
of singularity relative to the boundary of the singularity
\Arn.  Note that the coefficient of $y^2$ in \lines\ is
independent of $x$, and so all the lines meet at infinity.
This means that the difference of any two lines defines
a compact cycle of the singularity.  It is over these compact
cycles that we want to integrate $\Omega$.

With the change of variables above, $\Omega$ now takes the
form
\eqn\Omnew{\Omega ~=~ \bigg( {d \zeta \over \zeta}
\bigg)~\wedge~ {(~y + \beta^\prime(x) - x \alpha^\prime(x)~)~
dx \wedge dy  \over \sqrt{y^4  ~+~ b_1 (x,\zeta; u_j)~y^2 ~+~
b_0 (x,\zeta; u_j)}}}
At generic values of $x$ and $\zeta$ there are two branch cuts for
$\Omega$ (related by $y \to -y$) in the $y$-plane.  These cuts
simultaneously disappear at each of the $27$ lines.  As in \KLMVW,
the ``latitude circles'' of the homology 2-cycles correspond to
contours that surround the branch cuts, and these ``latitude
circles'' vanish at the North and South poles of the $2$-cycles
defined by the vanishing of the branch cuts at the lines.  In this
way, a pair of lines defines a homology $2$-cycle.  The only
difference here is that there are now two branch cuts (as opposed
to one such cut for $A_n$ and $D_n$), and this introduces a minor
subtlety.

Given two lines at which the cuts simultaneously vanish, there are
two ways in which the branch points can behave.  Suppose that near
one
line, $L$, the branch points appear in  pairs $\{y = \xi,
y = \eta \}$, and $\{y = - \xi, y = - \eta \}$  with $\xi \to \eta$
as one approaches $L$.  Take the cuts, $C_\pm$, to run from
$\pm \xi$ to $\pm \eta$ respectively.  As one approaches another
line,  $L^\prime$, either $\xi \to \eta$ again, or $\xi \to -\eta$.
For the former possibility, a contour surrounding either cut
defines a non-trivial $2$-cycle, and the two $2$-cycles
defined by the two cuts do not intersect.  If, however, one
has $\xi \to -\eta$, then the cuts $C_+$ and $C_-$ annihilate
one another.  Each cut then defines the half of a $2$-cycle,
and the annihilation of the cuts corresponds to gluing hemispheres
of the $2$-cycle together.  These two possibilities are
reflected in the fact that the difference of two different weights
in the $27$ of $E_6$ can either have length-squared $4$ or $2$.
For the former, the difference weights is the sum of two orthogonal
roots of $E_6$, and the latter corresponds to a single root of
$E_6$.  It follows that if the lines $L$ and $L^\prime$ give
rise to a pair of non-intersecting $2$-cycles, then there must me
another intermediate line, $L^{\prime \prime}$, such that
paths from $L$ to $L^{\prime \prime}$ and from
$L^{\prime \prime}$ to  $L^\prime$ describe each of the spheres
separately.

It should now be evident that the proper contour over which
to integrate $\Omega$ is one that surrounds both cuts, and that
does not get pinched off as  the cuts annihilate.
This simple closed loop in the $y$-plane can then
be deformed to large $|y|$, and the integral $\oint \Omega dy$
becomes the elementary integral: $\oint {dy \over y} = 2 \pi i$.
The integral over $x$ is now trivial: $2 \pi i\int dx$ evaluated
between the lines defining the poles and equatorial gluing of 
hemispheres  of the $2$-cycle in question. One thus finds that 
the integral of $\Omega$ over
the $2$-cycle, $L_{ij}$, defined by the lines $L_i$ and $L_j$ is:
\eqn\Omres{\int_{L_{ij}}\Omega ~=~ 2 \pi i~ (x_j ~-~ x_i)~
\bigg( {d \zeta \over \zeta}\bigg) ~=~ -\pi i~\big( \lambda_{SW}
\big|_{x = x_j} ~-~ \lambda_{SW} \big|_{x = x_i} \big) \ ,}
where $x_i$ and $x_j$ are the locations of the lines.  Thus
we recover the difference of the Seiberg-Witten differential
between the sheets of the Riemann surface, exactly as in \KLMVW.

Given that there are really only six compact $2$-cycles in the
fiber of the foliation, it follows that there are $12$ compact
$3$-cycles (six $A$-cycles and six $B$-cycles) in the total
space of the foliation \ALE.  Recalling the discussion of
the branch cuts in the base, and the plumbing of the Riemann
surface, we see that above each $\alpha_j$-cut, there are
six copies of the $\alpha_j$ $2$-cycle expanding from a
branch point and collapsing back to a branch point.  The result is
the $A$-type 3-cycle $\alpha_j$. (The $B$-type cycles are made by
going from
one of these branch points out to $\zeta = 0$ or $\zeta = \infty$.)
Thus the redundancy of the genus $34$ Riemann surface stems
from the highly redundant description of a six-dimensional
fiber homology in terms of $27$ lines.

This establishes a useful result about the integrals of
$\lambda_{SW}$ on the Riemann surface \Esixsurf: since there there
are only six $A$-type 3-cycles and six $B$-type 3-cycles, there can
only be six independent integrals of $\Omega$ over the sets of $A$
and $B$ cycles. This means that there can only be six independent
integrals of $\lambda_{SW}$ over the $A$-cycles of the Riemann
surface, and six independent integrals over $B$-cycles, even though
the genus of the surface is very large. Moreover, a basis of such
integrals can be obtained by choosing {\it any} representative cycle
in the surface that lies above each of the $\alpha_j$ cuts.

\newsec{Comments}

The lines on the surface in projective space, or the
non-compact cycles of the local singularity type, have played
a central role in defining the sheets of the Riemann surface.
This was also true of the analysis in \KLMVW, though it was not
explicitly stated there.  The fact that the lines in the
projective space span a vector space of  dimension $\ell + 1$
suggests that there may also be a further $2$-cycle or $3$-cycle
in  the Calabi-Yau manifold that is playing an interesting
implicit role.

While we have treated the $E_6$ theory in detail here, there is very
probably a similar story for the other $E_\ell$ groups. The
corresponding
singularities can all be realized as the blow-up of $\ell$ points in
general position ($\ell \le 8$) in a $\IP^2$ \Kanev. One can then
presumably realize these surfaces in some weighted projective spaces,
and then use the lines on these surfaces to relate the integration of
$\Omega$ over $3$-cycles to the integration of $\lambda_{SW}$ over
cycles of Riemann surfaces, in analogy to what we did above.

In \JMDN\ the $E_6$ singularity and the $27$ lines were used to
construct a candidate Seiberg-Witten curve for a superconformal
theory with matter and $E_6$ global symmetry. In string theory,
global symmetries are generically gauged \BDS, and so one might
expect some connection between the results here and those of \JMDN\
via some kind of flavour gauging. It turns out that we can come
fairly close: there are two extreme degenerations of \Esixsurf\ at
which the theory becomes superconformal: (i) the $E_6$
Argyres-Douglas points \doubref\AD\TEKH, and (ii) the point where the
Argyres-Douglas points come together: $\mu = 0$, {\it i.e.} when
$\tau = \zeta + u_6$. In the latter instance, the Riemann surface
collapses to a genus 4 surface foliated over the $x$-sphere:
\eqn\discrim{\bigg[~{x^3~(\zeta ~+~ u_6) ~-~ \
q_1\over p_1 } ~\bigg]^2 ~=~ p_2\ .}
The Seiberg-Witten differential, \lamSW, is holomorphic on this
surface except at infinity and at the zeroes of $\zeta$. At infinity
the differential has a double pole, while at the zeroes of $\zeta$ it
has a simple pole with residue $-2x$. From \Pasprod, \ptilde\ and
\Rsurf\ one immediately sees that there are $27$ such points on the
base $x$-sphere, and the residues are simply $-2 \lambda \cdot v$. If
one views the six-dimensional vector $v$ as the masses of some
matter fields,
then one has obtained a theory with matter with an $E_6$ global
symmetry. The most significant physical difference between this
matter theory, and that of \JMDN, is that this ``matter theory''
lacks an additional free parameter, called $\rho$ in \JMDN.
This parameter describes a Higgs v.e.v. on the Coulomb branch,
but in our degeneration limit, the Coulomb branch has been
collapsed to provide the masses.

In summary, we have made precise (for $E_6$) something that was
conjectured some while ago \KLTY: namely that the $ADE$ singularity
types must be the key ingredients in the construction of the quantum
effective actions of the $N=2$ supersymmetric Yang-Mills theories
with $ADE$ gauge groups. As in \KLMVW\ we have also established an
important string theory result: the $E_6$ Yang-Mills theory can be
represented in terms of compactifying the six-dimensional self-dual
string \sdstrings\ on the genus $34$ Riemann surface defined by
\Esixsurf.

\goodbreak
\vskip2.cm\centerline{{\bf Acknowledgements}}

We would like to thank the Aspen Center for Physics for providing a
stimulating research atmosphere and the opportunity to complete this
work. We would like to thank P.~Candelas, T.\ Eguchi, B.~Greene,
K.\ Hori, D.~Nemeschansky, and particularly E.~Martinec, S.~Katz and 
D.~Morrison for valuable discussions. Some of the details of the Riemann 
surface described in section 2 were done as part of a collaboration 
with E.~Martinec \MW, but were not published. 
N.W. is also grateful to the theory
division at CERN for hospitality while some of this work was
undertaken. N.W. is supported in part by funds provided by the DOE
under grant number DE-FG03-84ER-40168.

\goodbreak

\listrefs
\end